\documentstyle[12pt]{article}
\input epsf

\newlength{\dinwidth}
\newlength{\dinmargin}
\newcommand{\resection}[1]{\setcounter{equation}{0}\section{#1}}

\begin{document}
\vspace*{5cm}
\begin{center}
  \begin{Large}
  \begin{bf}
TRILINEAR ANOMALOUS GAUGE COUPLINGS AND NON-STANDARD FERMIONIC COUPLINGS\\
  \end{bf}
  \end{Large}
  \vspace{5mm}
  \begin{large}
R. Casalbuoni,  S. De Curtis \\
  \end{large}
Dipartimento di Fisica, Univ. di Firenze\\
I.N.F.N., Sezione di Firenze\\
  \vspace{5mm}
  \begin{large}
D. Guetta \\
  \end{large}
Dipartimento di Fisica, Univ. di Bologna  \\
I.N.F.N., Sezione di Bologna 
  \vspace{5mm}
  \vspace{5mm}
\end{center}
  \vspace{2cm}
\begin{center}
University of Florence - DFF 258/10/96 
\end{center}
\vspace{1cm}
\noindent
\newpage
\thispagestyle{empty}
\begin{quotation}
\vspace*{5cm}

\begin{center}
\begin{bf} 
  ABSTRACT
  \end{bf}
\end{center}
\vspace{1cm}\noindent
In this paper we do an analysis of the reaction $e^+e^-\to W^+W^-$ for various options of 
the Next Linear Colliders (NLC), by considering the possibility of deviations from the 
Standard Model induced by anomalous trilinear vector boson couplings and non-standard
fermionic couplings. We show that there are strong correlations among these parameters. By 
studying the high-energy behaviour of the helicity amplitudes we show that the error
made in linearizing the cross-section in the anomalous and non-standard couplings increases
with the square of the center of mass energy. We consider also the option of 
longitudinally polarized electron beams by showing that, in this case, the restrictions
on the anomalous and non-standard parameters are greatly enhanced.

\noindent

  \vspace{5mm}
\noindent
\end{quotation}
\newpage
\setcounter{page}{1}
\newcommand{\ac}{\`{a} \  }
\newcommand{\ec}{\`{e} \  }
\newcommand{\be}{\begin{equation}}
\newcommand{\ee}{\end{equation}}
\newcommand{\bea}{\begin{eqnarray}}
\newcommand{\eea}{\end{eqnarray}}  
\newcommand{\W}{\tilde{W} }  
\newcommand{\Y}{\tilde{Y} } 
\newcommand{\et}{\tilde{e} }       
\newcommand{\M}{\tilde{M} }    
\newcommand{\g}{\tilde{g} } 
\newcommand{\pli}{\tilde{\psi}_{L} }
\newcommand{\pri}{\tilde{\psi}_{R} }   
\newcommand{\bW}{{\bf \tilde{W}} }
 \newcommand{\bV}{{\bf V} }   
 \newcommand{\bY}{{\bf \tilde{Y}} }   
 \newcommand{\st}{\tilde{s}_{\theta} } 
 \newcommand{\ct}{\tilde{c}_{\theta} }
 \newcommand{\s}{s_{\theta} }
  \newcommand{\kt}{c_{\theta} } 
  \newcommand{\kdt}{c_{2\theta} }
 \newcommand{\A}{\tilde{A} }     
 \newcommand{\Z}{\tilde{Z} }     
 \newcommand{\q}{(\frac{g}{g^{\prime\prime}})^{2}}

  \newcommand{\epem}{e^{+}e^{-}} 
  \newcommand{\sd}{s_{2\theta} }    
  \newcommand{\te}{t_{\theta} }      
 \newcommand{\lnab}{ieg_{_{ZWW}}[g^{\nu\alpha}(-2p_{3}-p_{4})^{\beta}
 +g^{\alpha\beta} (p_{3}-p_{4})^{\nu}+g^{\beta\nu}(2p_{4}+p_{3})^{\alpha}]}    
 \newcommand{\abzq}{(a_{e}^{Z})^{2}+(b_{e}^{Z})^{2}}  
 \newcommand{\abz}{a_{e}^{Z}+b_{e}^{Z}}  
 \newcommand{\aez}{a_{e}^{Z}}  
 \newcommand{\bez}{b_{e}^{Z}}

\newcommand{\nn}{\nonumber}
\newcommand{\dd}{\displaystyle}
\newcommand{\AAA}{\frac{e^{2}}{4} \sin\phi}
\newcommand{\AAB}{\frac{e^{2}}{4\sqrt{2}} (\cos\phi-1)}
\newcommand{\AAC}{\frac{e^{2}}{4\sqrt{2}} (\cos\phi+1)} 
\resection{Introduction}

The sector of the Standard Model (SM) which has been directly verified
is the one of the
interactions of fermions with gauge vector bosons. However, the
non-abelian structure of the SM has been tested only in a very poor
manner and  indirectly through radiative corrections. The present and
future linear collider experiments will offer the opportunity to measure
the vector boson self-couplings at a high precision level \cite{murayama}. 
This will be
done via pair production processes of $W$, $Z$ and $\gamma$. The first
goal of such experiments will be a confirmation of the SM predictions. Besides,
they may be used to probe for new physics. In
particular it is  possible that signals for new physics will appear in
the gauge boson sector through the discovery of anomalous vertices.

In this paper we will  study the reaction $e^+e^-\to 
W^+W^-$ \cite{hagiwara}
with the $W$-pair decaying into a lepton pair plus jets. We will 
consider possible deviations from the SM predictions
 coming both from anomalous 
gauge vector boson trilinear couplings and from non-standard couplings  of the 
gauge   vector bosons to fermions. Also deviations of the $W$-mass from its SM 
value will be taken into account.
The reason for this study is that we can 
hardly imagine a scenario of new physics in which only the trilinear couplings  
are modified in a significant way from their SM values. As an example, let us 
consider the case of new physics leading to oblique corrections. The 
modifications induced in the $Z$ propagator will affect 
both the $Z$-coupling to a $W$-pair and the $Z$-coupling to a fermion pair, 
and we expect the deviations of the trilinear couplings  to be of the 
same order of magnitude of the deviations in the fermionic couplings. 
 In general the trilinear and the fermionic couplings result 
 strongly correlated (see later), therefore it is not justified to 
neglect the latter in a phenomenological analysis. Furthermore
we expect both these deviations to be small due to the LEP1 bounds on the 
fermionic couplings. The assumption of smallness 
of the trilinear anomalous couplings is also compatible with the fact that 
LEP1 has  tested  the SM at the level of radiative corrections.

In the present 
study we will consider also the experimental possibilities of having some 
polarization \cite{polarization} in the incoming electron beam and of reconstructing the 
polarization of the final $W$'s, 
because this should increase the sensitivity to the anomalous couplings \cite{paver}.
Theoretically, 
this program is carried out by 
evaluating the helicity amplitudes 
$F^{\lambda\lambda^\prime}_{\tau\tau^\prime}$, where 
$\lambda$ ($\lambda^\prime$) are the $e^-$ ($e^+$) helicities 
($\lambda^\prime=-\lambda=\pm 1/2$), and $\tau$ ($\tau^\prime$) = $\pm 1, 0$ 
are the $W^-$ ($W^+$) helicities.
In the calculation of the differential cross-sections we have not made
any approximation in the anomalous couplings. In fact, as it turns out
from the study of the high-energy behaviour, the quadratic terms in
these couplings are more and more important as the energy increases.

In Section 2 we will write the general parametrization of the gauge
boson trilinear anomalous couplings and of the non-standard fermionic
couplings. We will also evaluate the various helicity amplitudes. In
Section 3 we will analyze the high-energy behaviour of the
cross-sections relevant to our study and we will comment about the
expansion in the anomalous parameters. In Section 4 we will introduce
the observables considered in our analysis and we will discuss the
experimental errors. In Section 5 we will show the constraints on the
anomalous couplings from future $e^+e^-$ colliders. The options we
will consider are for  center of mass energies of 360, 500 and 800
$GeV$, with corresponding luminosities of 10, 20 and 50 ${\rm fb}^{-1}$.
Although a general fit could be possible we have limited our analysis to
the case of two parameters, by choosing one as an anomalous trilinear
coupling, and the other  proportional to non-standard
fermionic couplings. For completeness we will give in Appendix A the
high-energy expansion of the helicity amplitudes.

\resection{Helicity Amplitudes}

Let us now start the evaluation of the helicity amplitudes by defining the 
anomalous couplings we are interested in. We will consider only 
the $C$ and $P$ invariant possible trilinear couplings as defined by the 
lagrangian \cite{effective}
\bea 
{\cal L}^{3}_{eff} & = & 
   -ie[A_{\mu}(W^{-\mu\nu}W^{+}_{\nu} - W^{+\mu\nu}W^{-}_{\nu})+
   A^{\mu\nu}W^{+}_{\mu}W^{-}_{\nu} ]-
   iex_{\gamma }A^{\mu\nu}W^{+}_{\mu}W^{-}_{\nu} \nonumber \\
   &   & -ie(ctg\theta +\delta _{z} )
   [Z_{\mu}(W^{-\mu\nu}W^{+}_{\nu}-W^{+\mu\nu}W^{-}_{\nu})]-
   iex_{z} Z^{\mu\nu}W_{\mu}^{+}W_{\nu}^{-} \nonumber \\
   &   & +ie\frac{y_{\gamma}}{M_{W}^{2}}A^{\nu\lambda}W^{-}_{\lambda\mu}
   W^{+\mu}_{\nu} +ie\frac{y_{z}}{M_{W}^{2}}Z^{\nu\lambda }W^{-}_{\lambda\mu}
   W^{+\mu}_{\nu}  \label{lan} 
\eea
where $V_{\mu\nu}=\partial_\mu V_\nu-\partial_\nu V_\mu$ for $V=A,~Z,~W$.
As it is known, the parameters $x_\gamma$ and $x_z$ are related to the 
deviations of 
the dipole couplings from their SM values. The parameters $y_\gamma$ and $y_z$ 
give non-zero quadrupole interactions and $\delta_z$ is an overall deviation of 
the SM $ZW^+W^-$ coupling. 

We will assume 
that the energy relevant to the problem (i.e. the energy of the 
electron-positron beams) is much lower than the energy scale of the new 
physics. As a consequence we will work in the spirit of the effective 
lagrangians, by assuming all the anomalous couplings as constant and ignoring 
their possible energy dependence. 

The relevant charged and neutral couplings of $W$ and $Z$ to fermions are defined by 
the following lagrangians
 \be 
 {\cal L}^{neutral} =
  -eZ_{\mu}\bar{\psi}_e
  [\gamma^{\mu}a_{e}^{Z}-\gamma^{\mu}\gamma_{5}b_{e}^{Z}]\psi_e 
  -eA_{\mu}\bar{\psi}_e
  Q_{e}\psi_e\label{Lbn}  
  \ee
  \be  
  {\cal L}^{charged}= e  a_{w}\left(
 \sum_{\ell}\bar{\psi}_{\nu_{\ell}}\gamma^{\mu}(1-\gamma_{5})
  \psi_{\ell}W^{+}_{\mu} + 
  \sum_{i=1}^3\bar{\psi}_{i}\gamma^{\mu}(1-\gamma_{5})\tau_+
  \psi_{i}W^{+}_{\mu}+
  h.c. \right)
  \ee
In these expressions $\psi_e$ is the electron field, $\ell=e,\mu,\tau$, and
the sum over $i=1,2,3$ is on the quark charged current eigenstates. In the previous 
equations we will put the parameters $a_e^Z$, $b_e^Z$ and $a_W$ equal to 
their SM values plus anomalous terms, i.e. 
 \bea  
  a_{e}^{Z} & = & a_{e}^{Z}(SM) + \delta a_e^Z \nonumber \\
	   b_{e}^{Z} & = & b_{e}^{Z}(SM) + \delta b_e^Z \nonumber \\
	   a_{W} & = & a_{W}(SM) + \delta a_W 
 \eea
with
\be  a_{e}^{Z}(SM)=
\frac 1 {2\s\kt}(-\frac{1}{2}+2\s ^{2})
\ee
\be 
b_{e}^{Z}(SM) =  -\frac{1}{4\s\kt}
\ee
\be 
a_{W}(SM) = \frac{1}{2\sqrt{2}\s}  
\ee
where $\s$ is defined through the input parameters $G_F$, $\alpha(M_Z)$, $M_Z$
\be
\frac {G_F}{\sqrt{2}}=\frac{4\pi\alpha(M_Z)}{8\s^2\kt^2 M_Z^2}
\ee
or
\be
\s^2=\frac 1 2-\sqrt{\frac 1 4-\frac{\pi\alpha(M_Z)}{\sqrt{2}G_F M_Z^2}}
\ee
Finally we put
\be
M_W=M_W(SM)+\delta M_W
\ee
with
\be
M_W(SM)=\kt M_Z
\ee

The differential cross-section for initial $e^+_{\lambda^\prime}$, 
$e^-_\lambda$, and final $W^+_{\tau^\prime}$, $W^-_\tau$ is given by 
\be
       \frac{d\sigma ^{\lambda \lambda ^{\prime}}_{\tau \tau^{\prime}}}
	   {d\cos\phi}=\frac{|\vec{p}|}{4\pi s \sqrt{s}}
	   |F^{\lambda\lambda^{\prime}}_{\tau\tau^{\prime}}(s,\cos\phi)|
	   ^{2}
\ee
where
\be
|\vec p|=\beta_W\frac{\sqrt{s}} 2,~~~~~\beta_W=\sqrt{1-\frac{4M_W^2}{ s}}
\ee
and $\phi$ is the angle between the incoming electron and $W^-$.

To express the results for the helicity amplitudes
we will use the same formulation as given in ref. 
\cite{LEP2}. We start separating  the contributions from photon, $Z$, and neutrino 
\be
F_{\tau\tau^\prime}^{\lambda\lambda^\prime}=F=F_\gamma+F_Z+F_\nu
\ee
Then we define reduced amplitudes $\tilde F$ by factorizing out the leading 
angular dependence in terms of the rotation functions $d^J_{\mu,\mu^\prime}$.
Here $J=1,2$ corresponds to the lowest angular momentum which contributes to a 
given helicity amplitude. We will write
\be
F_{\tau\tau^\prime}^{\lambda\lambda^\prime}=\sqrt{2}\lambda e^2
d^J_{2\lambda,\tau-\tau^\prime}{\tilde 
F}_{\tau\tau^\prime}^{\lambda\lambda^\prime}
\ee
We follow here the phase conventions as in Gounaris et al. in ref. \cite{LEP2}. 
However, our helicity amplitudes  $F$ correspond to their ${\cal M}/2$.
In the case of transverse final $W$'s, and for  $(\tau,\tau^\prime)=(\pm 1,\mp 
1)$, corresponding to  $J=2$, the electron must be 
left-handed, and the helicity amplitude takes contribution only from the 
neutrino exchange. We obtain
\be
F^{-1/2,1/2}_{\tau,-\tau}=-4\tau e^2\left(\frac 1{2\sqrt{2}\s}+\delta 
a_W\right)^2\frac s t \sin\phi\left(\frac{1-\tau\cos\phi} 2\right)
\ee
The other helicity amplitudes involve all the contributions and their reduced 
amplitudes can be expressed in the following way (separating the various 
contributions)
\be
{\tilde F}_\gamma=-\beta_W A^\gamma
\ee
\bea
{\tilde F}_Z &=&\beta_W A^Z\Big(\delta_{\lambda,1/2}(1+\cot\theta (\delta 
a_e^Z-\delta b_e^Z))\nn\\
&&+ \delta_{\lambda,-1/2}(1-\frac 1{2\s^2}+\cot\theta (\delta 
a_e^Z+\delta b_e^Z)) \Big)\frac s {s-M_Z^2}
\eea
\be
{\tilde F}_\nu=\delta_{\lambda,-1/2}\frac{4}{\beta_W}\left(\frac 
1{2\sqrt{2}\s}+\delta a_W\right)^2\left(B_{\tau\tau^\prime}+\frac 
s{4t}C_{\tau\tau^\prime}\right)
\ee
where the Mandelstam variable $t$ is defined by
\be 
t=M_W^2-\frac s 2 (1-\beta_W\cos\phi)
\ee
and the coefficients $A^V$, $B$ and $C$ are given in Table 1.
\begin{table} 
\caption{Reduced helicity amplitudes 
${\tilde F}_{\tau\tau^\prime}^{\lambda\lambda^\prime}$,
for the process $ e^+e^-\to W^+W^- $.}
%
\vspace{0.4cm}
\begin{center}
\begin{tabular}{|c||c|c|c|c|}
\hline
&&&&\\
$(\tau,\tau^\prime)$&$A^V_{\tau,\tau^\prime}$ &$B_{\tau,\tau^\prime}$ & 
$C_{\tau,\tau^\prime}$& $d^J_{2\lambda,\tau-\tau^\prime}$\\
& & & & \\ \hline
&&&&\\
$(+1,+1)$ & $g_1^V+\dd{\frac s{2M_W^2}} \lambda_V$ & 1 &  
$\dd{\frac {4 M_W^2}{s }}  $ & $-\dd{\frac{2\lambda}{\sqrt{2}}}\sin\phi $\\ 
& & & & \\ \hline
&&&&\\
$(-1,-1)$& $g_1^V+\dd{\frac s{2M_W^2}} \lambda_V$  & 1 &  
$\dd{\frac {4 M_W^2}{s }} $ &
$-\dd{\frac{2\lambda}{\sqrt{2}}} \sin\phi  $\\
& & & & \\ \hline
&&&&\\
$(+1,0)$ & $\dd{\frac {\sqrt{s}}{2M_W}} (g_1^V+\kappa_V+\lambda_V)$ &  $ 
\dd{\frac {\sqrt{s}}{M_W}}  $& $ \dd{\frac{4M_W}{\sqrt{s}}}(1+\beta_W) $ & 
$\dd{\frac{(1+2\lambda \cos\phi)}2}$\\
& & & & \\ \hline
&&&&\\
$(0,-1)$ & $\dd{\frac {\sqrt{s}}{2M_W}}  (g_1^V+\kappa_V+\lambda_V)$ &  $ 
\dd{\frac {\sqrt{s}}{M_W}}  $& $ \dd{\frac{4M_W}{\sqrt{s}}} (1+\beta_W) $ & 
$\dd{\frac{(1+2\lambda \cos\phi)}2} $\\
& & & & \\ \hline
&&&&\\
$(0,+1)$ & $\dd{\frac {\sqrt{s}}{2M_W}}  (g_1^V+\kappa_V+\lambda_V)$ &  $ 
\dd{\frac {\sqrt{s}}{M_W}}  $& $ \dd{\frac{4M_W}{\sqrt{s}}}(1-\beta_W)$ & 
$\dd{\frac{(1-2\lambda \cos\phi)}2} $\\
& & & & \\ \hline
&&&&\\
$(-1,0)$ & $\dd{\frac {\sqrt{s}}{2M_W}} (g_1^V+\kappa_V+\lambda_V)$ &  $ 
\dd{\frac {\sqrt{s}}{M_W}}  $& $\dd{\frac{4M_W}{\sqrt{s}}} (1-\beta_W)$ & 
$\dd{\frac{(1-2\lambda \cos\phi)}2} $\\
& & & & \\ \hline
&&&&\\
$(0,0) $& $ g_1^V+\dd{\frac s{2M_W^2}}\kappa_V $ &  $ \dd{\frac { s}{2M_W^2}} 
$& $ \dd{\frac {8 M_W^2} {s }} $ & $ -\dd{\frac{2\lambda}{\sqrt{2}}}  
\sin\phi$\\
& & & & \\ \hline
\end{tabular}
\end{center}
\end{table} 
Here, $V=\gamma,~Z$, and we have used the following definitions
\be
g_1^\gamma=1,~~~~g_1^Z=1+\delta_z\tan\theta
\ee
\be
\kappa_\gamma=1+x_\gamma,~~~~\kappa_Z=1+(x_z+\delta_z)\tan\theta
\ee
\be
\lambda_\gamma=y_\gamma,~~~~~\lambda_Z=y_z\tan\theta
\ee

 It is worth to notice that the helicity amplitudes for two longitudinally 
 polarized $W$'s do not depend on the quadrupole couplings $y_z$ and 
$y_\gamma$.

\resection{High-energy behaviour}

In order to have a qualitative idea of the behaviour of the helicity
amplitudes, 
 we will consider here a double expansion  in powers
of the energy and of  the anomalous couplings. In the phenomenological
analysis, however, we will use the exact expressions.

Within the SM the $e^+e^-\to W^+W^-$ cross-section has a good behaviour 
at high-energy 
due to the cancellations required by the gauge-invariance. In the actual 
case, however,  this is not so. We should remember that we are dealing with 
an effective theory and that we are allowed to use it only up to energies 
smaller 
with respect to the new physics scale. The analysis of the high-energy 
behaviour is particularly simple in terms of the helicity amplitudes, in fact 
the cancellations operating at the level of the SM must be effective within 
each single helicity amplitude (remember that the cross-section is a sum of 
modulus squares of helicity amplitudes).   The leading behaviour in energy 
is simply evaluated  just counting the powers of energy associated 
to the various vertices, propagators and external wave functions. The worst 
behaved amplitudes are the ones corresponding to the production of two 
longitudinal vector bosons. The power counting would imply that the corresponding 
amplitudes (let us call them shortly $F_{LL}$) should increase as $s^2$, due to 
the fact that the vertices coming from the quadrupole interactions contain 
three derivatives. This is not the case since, as already noticed, 
these amplitudes do not depend on the couplings $y_z$ and $y_\gamma$ signalling 
that the quadrupole vertices do not contribute to $F_{LL}$. In fact we 
can show that, due to the structure of the quadrupole interaction, the 
corresponding high-energy behaviour of the helicity amplitudes is not the one 
coming from simple power counting but it is suppressed by one power of $s$. The 
quadrupole interactions are made up entirely in terms of the abelian field 
strengths $A_{\mu\nu}$, $Z_{\mu\nu}$ and $W_{\mu\nu}$, therefore if we 
decompose the helicity amplitudes in the form
\be 
F_{\tau\tau^\prime}^{\lambda\lambda^\prime}=
F_{\mu\mu^\prime}^{\lambda\lambda^\prime} \epsilon^\mu_\tau(p^-)
\epsilon^{\mu^\prime}_{\tau^\prime}(p^+)
\ee
where $p^\pm$ are the momenta of $W^\pm$ and 
$\epsilon^\mu_\tau$ is the polarization vector for the $W$'s,  
the quadrupole parts of the amplitudes must satisfy
\be
(p^-)^\mu F_{\mu\mu^\prime}^{\lambda\lambda^\prime}({\rm quadr.})=
(p^+)^{\mu^\prime} F_{\mu\mu^\prime}^{\lambda\lambda^\prime}({\rm quadr.})=0
\ee
In the high-energy limit the longitudinal polarization vectors have the 
following behaviour
\be
\epsilon^\mu_L(p)=\frac{p^\mu}{M_W}\left(1+{\cal O}\left(\frac 1 
s\right)\right)
\ee
It follows that $y_z$ and $y_\gamma$ cannot give contribution to the highest 
possible power in $s$ in amplitudes relative to longitudinally polarized $W$'s.
As a consequence the highest  power for $F_{LL}$ is $s$ instead of 
$s^2$, whereas for $F_{TL}$ it is $s^{1/2}$  instead of $s^{3/2}$.

Here we will consider the corrections to the helicity amplitudes up to the first
order in the anomalous couplings and in $\delta M_W$ (see eq. (2.10))
\be
F_{\tau\tau^\prime}^{\lambda\lambda^\prime}=
{F}_{\tau\tau^\prime}^{\lambda\lambda^\prime}(SM)+
\delta F_{\tau\tau^\prime}^{\lambda\lambda^\prime}
\ee
where ${F}_{\tau\tau^\prime}^{\lambda\lambda^\prime}(SM)$ 
are the SM helicity amplitudes, and  then we will
expand the generic helicity amplitude, $F$, in powers of energy
\be
F=\sum_\alpha F^{(\alpha)} s^\alpha
\ee
One gets, for the longitudinal-longitudinal case
\be
F_{LL}=\delta F_{LL}^{(1)} s+F_{LL}^{(0)}(SM)+\delta F_{LL}^{(0)}+\frac 1 s 
\left(F_{LL}^{(-1)}(SM)+\delta F_{LL}^{(-1)}\right)+
{\cal O}\left(\frac 1 {s^2}\right) 
\ee
for the longitudinal-transverse case 
\be
F_{LT}=\sqrt{s}\left(\delta F_{LT}^{(1/2)}+\frac 1 s \left(F_{LT}^{(-1/2)}(SM)+ 
\delta F_{LT}^{(-1/2)}\right)+{\cal O}\left(\frac 1 {s^2}\right) \right)
\ee
and finally the expansion for $F_{TT}$ is
\be
F_{TT}=\delta F_{TT}^{(1)} s+F_{TT}^{(0)}(SM)+\delta F_{TT}^{(0)}+\frac 1 s 
\left(F_{TT}^{(-1)}(SM)+\delta F_{TT}^{(-1)}\right)+
{\cal O}\left(\frac 1 {s^2}\right) 
\ee
Notice that  $F_{TT}$ has only contributions from $y_z$ and 
$y_\gamma$, because in this case they are not suppressed and, by power counting, 
the other anomalous contributions should give a constant term for $s\to\infty$.
The list of the coefficients $F^{(\alpha)}$ is given in Appendix A. 

We can now discuss the anomalous contributions in the energy expansion of the 
differential cross-sections.  
We get (we 
do not expand the threshold factor $\beta_W$, see eqs. (2.12) and (2.13));

\bea
\frac{d\sigma_{LL}}{d\cos\phi}& = &
\frac{\beta_W}{8\pi}\Big [|\delta F_{LL}^{(1)}|^2 s+
(\delta F_{LL}^{(1)})^* 
F_{LL}^{(0)}(SM)+\delta F_{LL}^{(1)} (F_{LL}^{(0)}(SM))^* \nn\\
& + &
\frac 1 s \sum_{\alpha=1,0}(\delta F_{LL}^{(\alpha)})^* 
F_{LL}^{(-\alpha)}(SM)+\delta F_{LL}^{(\alpha)} 
(F_{LL}^{(-\alpha)}(SM))^*\Big]+
{\cal 
O}\left(\frac 1 {s^2}\right)
\eea
\bea
\frac{d\sigma_{LT}}{d\cos\phi}&=&
\frac{\beta_W}{8\pi} \Big[|\delta F_{LT}^{(1)}|^2 +
\frac 1 s\Big((\delta F_{LT}^{(1/2)})^* 
F_{LT}^{(-1/2)}(SM)+\delta F_{LT}^{(1/2)} (F_{LT}^{(-1/2)}(SM))^*\Big) 
\Big]\nn\\
&+&{\cal O}\left(\frac 1 {s^2}\right)
\eea
\bea
\frac{d\sigma_{TT}}{d\cos\phi}&=&
\frac{\beta_W}{8\pi}\Big [|\delta F_{TT}^{(1)}|^2 s+
(\delta F_{TT}^{(1)})^* 
F_{TT}^{(0)}(SM)+\delta F_{TT}^{(1)} (F_{TT}^{(0)}(SM))^*\nn\\
& +&
\frac 1 s \sum_{\alpha=1,0}(\delta F_{TT}^{(\alpha)})^* 
F_{TT}^{(-\alpha)}(SM)+\delta F_{TT}^{(\alpha)} 
(F_{TT}^{(-\alpha)}(SM))^*\Big]+
{\cal 
O}\left(\frac 1 {s^2}\right)
\eea
At each order in $s$ we have taken only the first non-zero terms
in the anomalous couplings. In particular the terms linear in $s$ turn out to be 
quadratic in the anomalous couplings.
Notice  that the true expansion parameter is $s/M_W^2$, 
therefore it is a bad approximation to neglect the quadratic terms in the 
anomalous parameters when one is discussing the cross-sections at energies of 
the order $10~M_W$ or more (we are assuming a size for the anomalous parameters 
comparable to the size of the radiative corrections within the SM).

As we have already observed $d\sigma_{LL}$ does not depend on $y_z$ and 
$y_\gamma$. Furthermore, by looking at the explicit expressions in Appendix A, 
we see that the most divergent terms in $d\sigma_{LT}$  
depend on $x_\gamma+y_\gamma$, $2\delta_z+x_z+y_z$, $\delta a_e^Z$, $\delta 
b_e^Z$, $\delta a_W$ and, in $d\sigma_{TT}$ depend only on
 $y_\gamma$, $y_z$. 
It should also be 
noted that $\delta M_W$ appears only at lower orders in the expansion. 
This 
study suggests that an analysis  of the differential 
cross-sections at various  energies should be useful in order to discriminate 
among 
various anomalous parameters.

\resection{Phenomenological analysis}

In the following phenomenological analysis we will consider  
differential cross-sections for longitudinally polarized electron and 
unpolarized
positron beams 
\be
\frac{d\sigma_{\tau\tau^\prime}}{d\cos\phi}=
\frac 1 4 \left[(1+P_e)\frac{d\sigma^{(+1/2,-1/2)}_{\tau\tau^\prime}}{d\cos\phi}+
(1-P_e)\frac{d\sigma^{(-1/2,+1/2)}_{\tau\tau^\prime}}{d\cos\phi}\right]
\ee
where $P_e$ is the degree of longitudinal polarization. 
For $P_e=0$ we get the expression for unpolarized electrons.
We have also considered the production of
pairs of polarized $W$'s. 

In order to study the 
sensitivity of the differential cross-sections to the anomalous gauge 
couplings and to the non-standard fermionic couplings, we divide the
experimentally significant range of $\cos\phi$ ($-0.95\le\cos\phi\le 0.95$)
into 6 bins. For each differential cross-section we can evaluate the 
differences
\be
\sigma_{i,a}^{AN}=\sigma_{i,a}-\sigma_{i,a}(SM),~~~~i=1,\cdots,6,~~~a=
LL,LT,TT
\ee
where $\sigma_{i,a}$ and $\sigma_{i,a}(SM)$ are respectively the full 
and the SM differential cross-sections integrated over the bin $i$. From 
these differences we can evaluate the $\chi^2$ function for each differential
cross-section
\be
\chi^2_a=\sum_{i=1}^6\frac{(\sigma_{i,a}^{AN})^2}{(\delta\sigma_{i,a})^2},
~~~~~~a=LL,LT,TT
\ee
where we assume the errors $\delta\sigma_{i,a}$ as follows
\be
\delta\sigma_{i,a}=\sqrt{\frac{\sigma_{i,a}}{BR\cdot L}+
\sigma_{i,a}^2 \Delta_{\rm sys}^2}
\ee
where $\Delta_{\rm sys}$ is the systematic error taking into account the relative errors
on the luminosity, on 
the branching ratio of the decay of $W$'s into a lepton pair plus jets,
on the acceptance and on the longitudinal polarization of the electron beam. We assume
 $\Delta_{\rm sys}=1.5\%$. We take into account the
efficiency in reconstructing the $W$ pairs from their decays into a lepton pair
plus jets by reducing the true branching ratio from 0.29 to 
0.10 (see \cite{BR}). This is the effective branching ratio $BR$ that we use in eq. (4.4).
There is a delicate point about the branching ratio. In fact we are assuming that 
the coupling $We\nu$  may differ from its SM value. This could modify the
branching ratios of $W$'s into fermions. However, for
deviation respecting the universality, it is easy to see that neglecting, as 
usual, mixed terms in the radiative and in the anomalous corrections, the
branching ratios are the same as in the SM up to 1-loop level.

\resection{Bounds on the anomalous couplings}

In this Section we will consider the bounds on the anomalous couplings coming from 
the minimization of the $\chi^2$-function as defined in the previous
Section.  We will not attempt to make a general analysis in all the
parameters expressing
the deviations from the SM (5 anomalous trilinear couplings,  3 non-standard
fermionic couplings and the $M_W$ deviation), but rather we will
concentrate on a few models depending only on 2 parameters. 
The first simplifying hypothesis we will
make is that all the fermionic deviations can be expressed in terms of the
$\epsilon_i$, $i=1,2,3$, parameters \cite{altarelli}. By using the definitions of 
the $\epsilon_i$ in terms of the observables one gets
\bea
\delta a_e^Z&=&-\frac {\epsilon_1}{8\s\kt}-\frac 1
{\kdt}\frac{\s}{\kt}\left(\frac{\epsilon_1} 2-\epsilon_3\right)\nn\\
\delta b_e^Z&=&\frac {\epsilon_1}{8\s\kt}\nn\\
\delta a_W&=&\frac 1{2\sqrt{2}\s}\left(\frac{\kt^2}{2
\kdt}\epsilon_1-\frac{\epsilon_2} 2-\frac{\s^2}{\kdt}\epsilon_3\right)
\nn\\
\frac{\delta
M_W^2}{M_W^2}&=&\frac{\kt^2}{\kdt}\epsilon_1-\epsilon_2-2\frac{\s^2}{\kdt}\epsilon_3 
\eea
In many models of new physics there is no isospin violation. In such a
case one has $\epsilon_1=\epsilon_2=0$. 
We will restrict our analysis to these models, therefore all the non-standard
fermionic couplings and $\delta M_W$ are parametrized in terms of $\epsilon_3$.
The experimental
bound on $\epsilon_3$ coming from LEP1 measurements ref.
\cite{HEPconf},  is
\be
\epsilon_3^{\rm exp}=(4.6\pm 1.5)\times 10^{-3}
\ee
Notice that the $\epsilon_3$ appearing in our equations is only the
contribution due to new physics, so in order to get an useful bound for
it one has to subtract from $\epsilon_3^{\rm exp}$ the contribution from
the SM radiative corrections. We will assume in our analysis
$m_{top}=175~GeV$ and a value of $m_{Higgs}$ ranging from 65 $GeV$ to 1
$TeV$. The corresponding values due to  radiative corrections are
\bea
\epsilon_3^{\rm rad}&=&4.65\times 10^{-3},~~~~~~m_{Higgs}= 65~GeV\nn\\
\epsilon_3^{\rm rad}&=&6.53\times 10^{-3},~~~~~~m_{Higgs}= ~~1~TeV
\eea
The 90\% C.L. bound we get (in the case of two degrees of freedom) is then
\be
-(4.5_{+0.6}^{-1.3})\times
10^{-3}\le\epsilon_3\le(1.9_{-0.6}^{+1.3})\times 10^{-3}
\ee
The upper and lower bounds correspond to $m_{Higgs}=65~GeV$ and
$m_{Higgs}=1~TeV$ respectively. The central value is obtained for
$m_{Higgs}= 300~GeV$, with a corresponding radiative contribution given
by
\be
\epsilon_3^{\rm rad}=5.91\times 10^{-3},~~~~~~m_{Higgs}= ~300~GeV
\ee
 
A comparable bound on $\epsilon_3$ will be obtained by the precision measure
of $M_W$ at LEP2. Assuming $\delta M_W=50~MeV$ we get from eq. (5.1), always
in the hypothesis of $\epsilon_1=\epsilon_2=0$,
\be
-3.2\times 10^{-3}\le\epsilon_3\le 3.2\times 10^{-3}
\ee

As far as the anomalous trilinear gauge couplings are concerned we will also consider some
simplified models.\\\\
\noindent
\underbar{MODEL 1}\\\\
A first possibility discussed in ref. \cite{bilenky} is
to assume a global $SU(2)_L$ symmetry for the lagrangian (2.1). This
gives rise to the following relations among the trilinear anomalous
couplings:
\be
\delta_z=x_\gamma=x_z=0
\ee
and
\be
y_z=\frac{\kt}{\s}y_\gamma\not=0
\ee
In this way, model 1 depends only on the two parameters $(\epsilon_3,y_\gamma)$.
We analyze the model for various observables, namely
the  unpolarized cross-sections for different final $W$'s polarization,
$\sigma_{LL}$, $\sigma_{LT}$, $\sigma_{TT}$ and the total one. The 90\%
C.L. bounds are
given in  Fig. 1, together with the bounds coming from combining the
$\chi^2$ for the final state polarized observables, $\sigma_{\rm comb}$. 
In the figure we give the bounds for
a NLC machine with a center of mass energy of 500 $GeV$ and a luminosity
of 20 $fb^{-1}$. The figure shows what we have already discussed before,
that is the independence of $\sigma_{LL}$ on $y_\gamma$. In this case
it is important to measure other differential cross-sections. Also the
total one, if combined with $\sigma_{LL}$ would give a good restriction
in the parameter space. In Fig. 2 we give the 90 \% C.L. bounds arising from
combining the final state polarized differential cross-sections at
various energies and luminosities, namely: $E=360~GeV,L=10~fb^{-1}$;
$E=500~GeV,L=20~fb^{-1}$; $E=800~GeV,L=50~fb^{-1}$. We see that one
needs  to reach at least 800 $GeV$ in energy in order to get bounds 
on $\epsilon_3$
comparable with the ones obtained at LEP1
also shown in the figure. The corresponding strip is 
for $65\le m_H(GeV)\le 1000$. Notice that the bound at $E=800~GeV$
includes also a region which is not connected with the SM point. This means that 
there is the possibility of having an ambiguity at this energy. However this is already
resolved by the LEP1 data.
It is also interesting to
notice that there are strong correlations among $y_\gamma$ and
$\epsilon_3$. This means that going at values of $\epsilon_3$ different
from zero, the bounds on $y_\gamma$ can change considerably. As we
will see this is a rather general feature.
\begin{figure}
\epsfxsize=8truecm
\centerline{\epsffile[50 263 506 693]{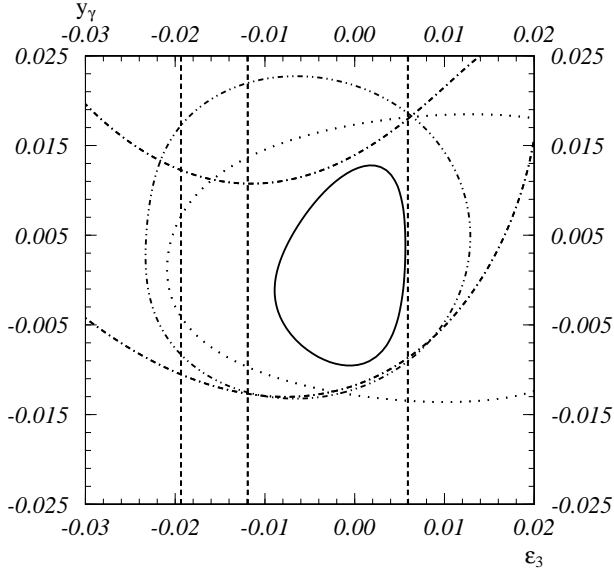}}
\caption{
90\% C.L. bounds on the parameter space $(\epsilon_3,y_\gamma)$
in the case of model 1. The dotted line represents the bounds from
$\sigma_{TT}$, the dash-dotted one from
$\sigma_{LT}$, the dashed one  from
$\sigma_{LL}$, the dash-doubledotted one  from
$\sigma_{\rm total}$, and the continuous one  from
$\sigma_{\rm comb}$.
Here the parameters of the NLC  
are $E=500~GeV$,  $L=20~fb^{-1}$, and $P_e=0$.}
\end{figure}
 \begin{figure}
\epsfxsize=8truecm
\centerline{\epsffile[50 263 506 693]{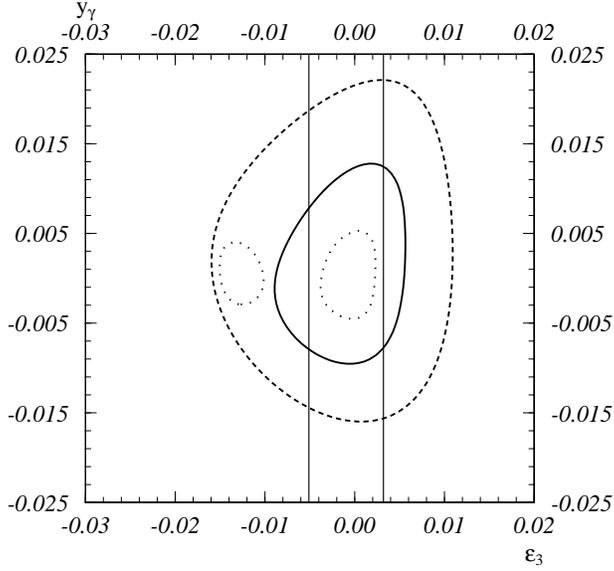}}
\caption{90\% C.L. bounds on the parameter space $(\epsilon_3,y_\gamma)$
in the case of model 1. The bounds are obtained from 
$\sigma_{\rm comb}$ at various
energies and luminosities. The dashed line corresponds to $E=360~GeV$
and $L=10~fb^{-1}$, the continuous one to $E=500~GeV$
and $L=20~fb^{-1}$, and the dotted one  to $E=800~GeV$,
$L=50~fb^{-1}$, and $P_e=0$ The vertical lines correspond to
the LEP1 data.}
\end{figure}
\begin{figure} 
\epsfxsize=8truecm
\centerline{\epsffile[50 263 506 693]{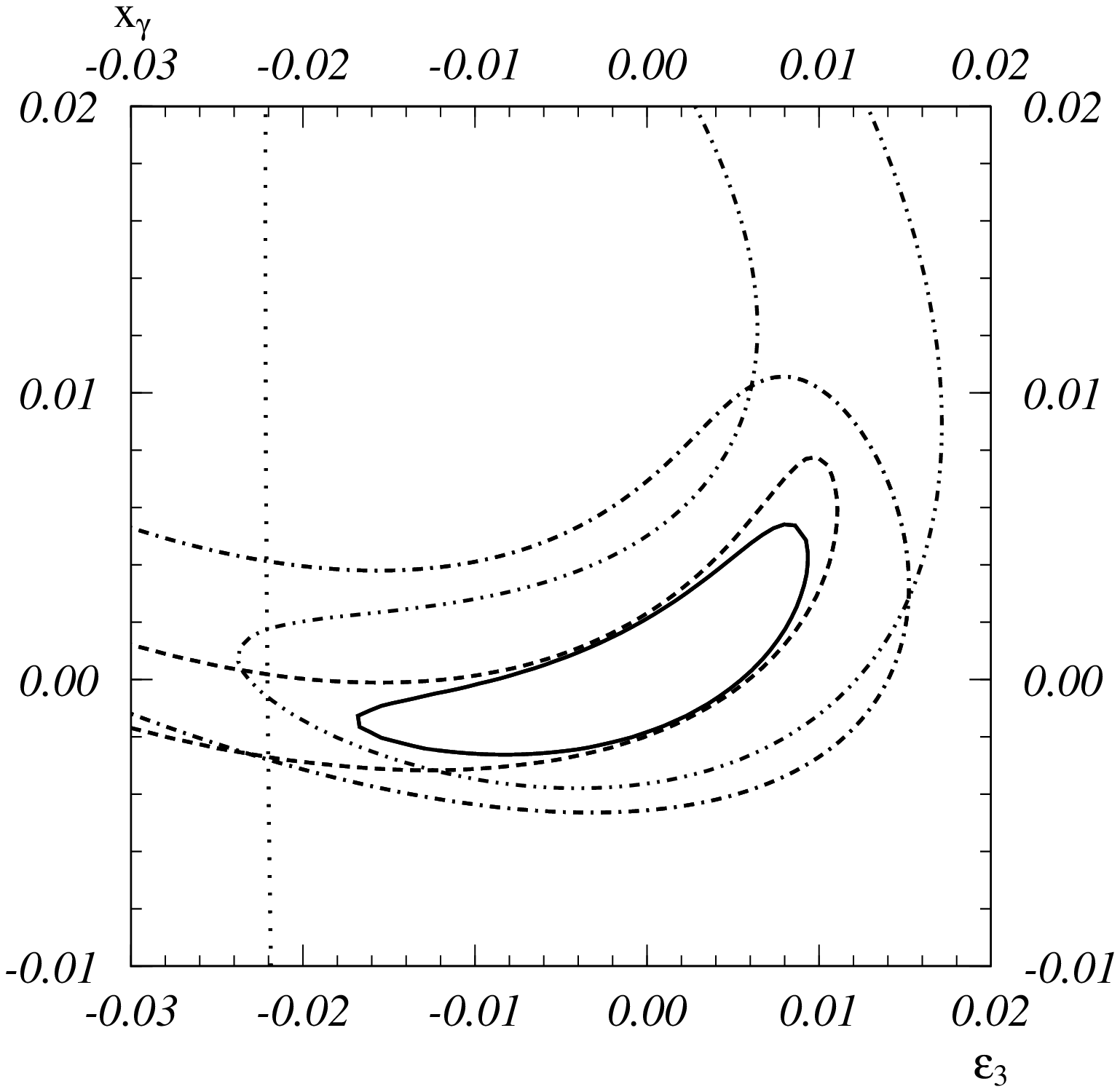}}
\caption{Same of Figure 1 for the parameter space $(\epsilon_3,x_\gamma)$
in the case of model 2.}
\end{figure}
\begin{figure} 
\epsfxsize=8truecm
\centerline{\epsffile[50 263 506 693]{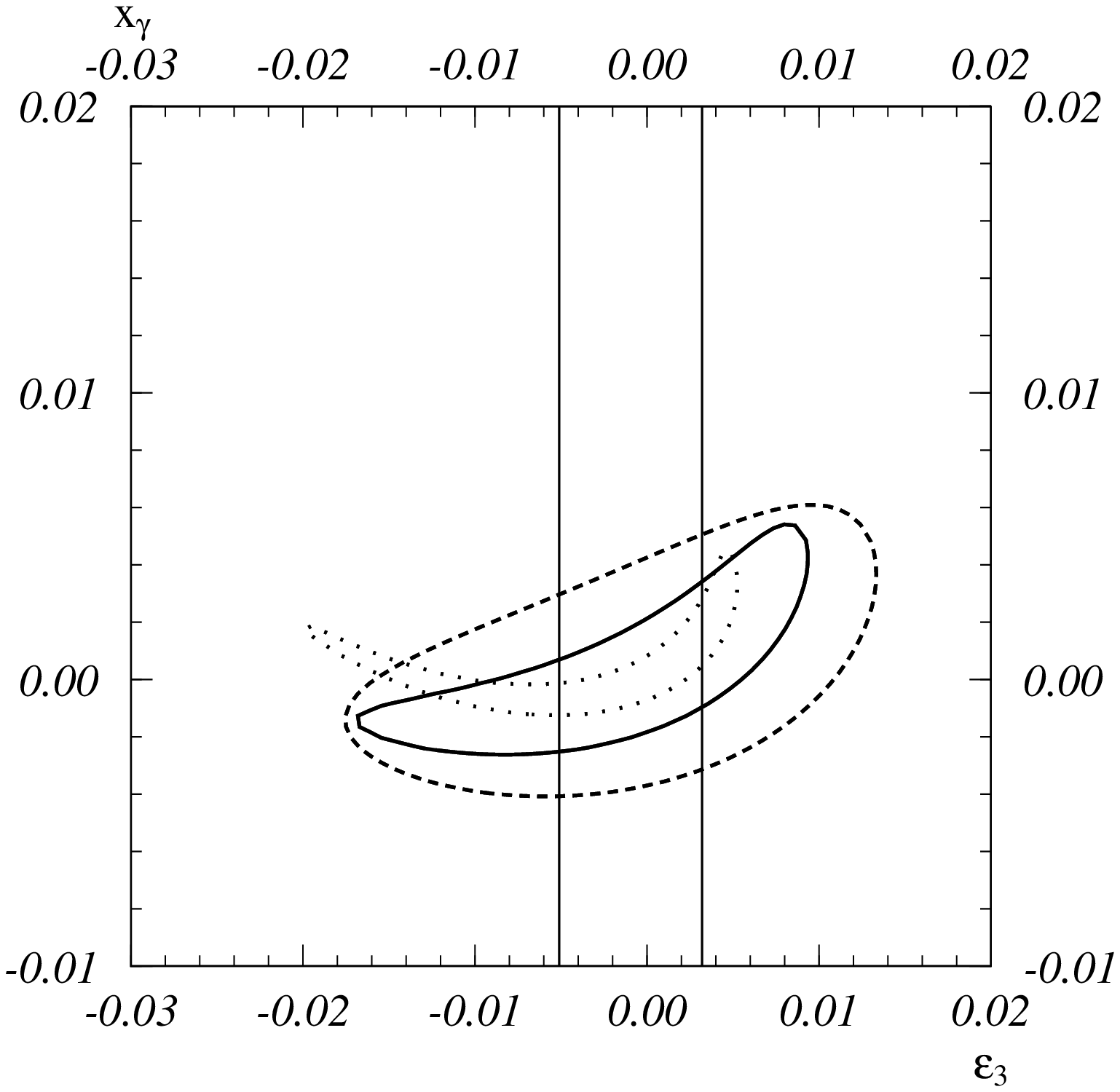}}
\caption{Same of Figure 2 for the parameter space $(\epsilon_3,x_\gamma)$
in the case of model 2.}
\end{figure}\\\\
\noindent\underline{MODEL 2}\\\\
Next we consider the case in which one adds to the SM lagrangian the
most general operator of dimension 6 invariant under $SU(2)_L\otimes
U(1)$ (see ref. \cite{bilenky}). The general form of this term is
\be
\frac{ie}{2M_W^2}\left[\frac{f_B}{\kt} B_{\mu\nu}
\left(D_\mu\phi\right)^\dagger\left(D_\nu\phi\right)+
\frac{f_w}{\s} {\vec w}_{\mu\nu}\cdot
\left(D_\mu\phi\right)^\dagger\vec\tau\left(D_\nu\phi\right)\right]
\ee
where $B_{\mu\nu}$ is the weak hypercharge field strength, $\vec
w_{\mu\nu}$ is the non-abelian field strength associated to ${\vec
W}_\mu$ and $D_\mu$ the covariant derivative operating on the Higgs field
$\phi$. The interaction generated by this operator is
non-renormalizable but, assuming $f_B=f_w$, the divergences soften and
only the logarithmic ones remain in the final result, whereas the
potential quadratic divergences are replaced by quadratic dependence
on the Higgs mass. The operator (5.9) induces anomalous couplings which,
with the further constraint $f_B=f_w$, satisfy the following relations
\be
x_z=-\frac{\s}{\kt}x_\gamma,~~~\delta_z=\frac 2{\s\kt} x_\gamma,~~~
y_\gamma=y_z=0 
\ee
Assuming again $\epsilon_3\not=0$ we get a two parameter space
$(\epsilon_3,x_\gamma)$. Fig. 3 is the analogous of Fig. 1 for the
present model. We notice that $\sigma_{LL}$ bounds strongly $x_\gamma$
but has a weaker influence on $\epsilon_3$. Also $\sigma_{TT}$ does not
depend on $x_\gamma$. In Fig. 4 we give again the bounds for various
energies and luminosities.  Finally let us
stress once more the strong correlation among the trilinear anomalous
coupling and $\epsilon_3$.
\\\\
\begin{figure} 
\epsfxsize=8truecm
\centerline{\epsffile[50 263 506 693]{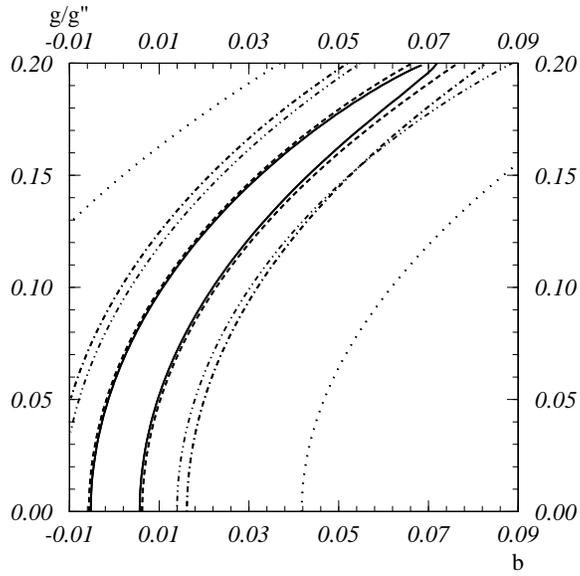}}
\caption{Same of Figure 1 for the parameter space $(b, g/g^{\prime\prime})$
in the case of model 3 (BESS model).}
\end{figure}
\begin{figure} 
\epsfxsize=8truecm
\centerline{\epsffile[50 263 506 693]{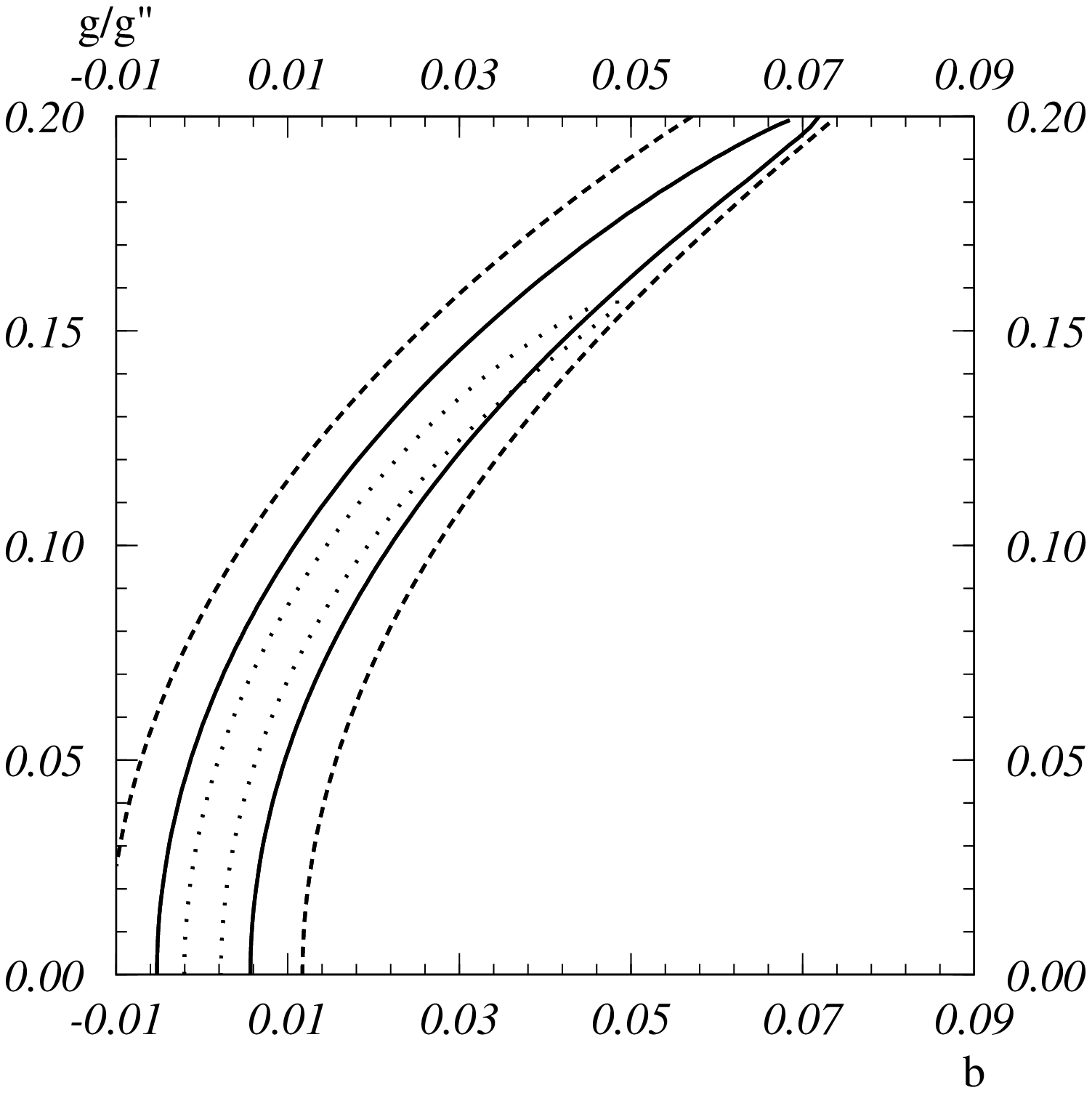}}
\caption{Same of Figure 2 for the parameter space $(b, g/g^{\prime\prime})$
in the case of  model 3 (BESS model).}
\end{figure}
\begin{figure} 
\epsfxsize=8truecm
\centerline{\epsffile[50 263 506 693]{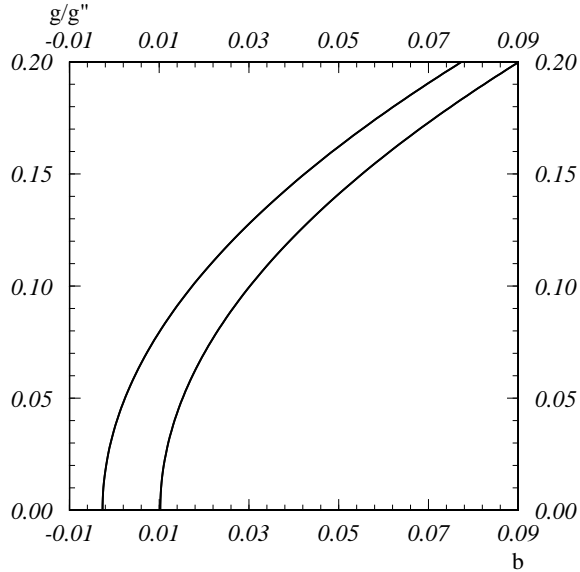}}
\caption{90\% C.L. bounds on the parameter space $(b,
g/g^{\prime\prime})$ in the case of model 3 (BESS model), as
obtained from LEP1 experiments.}
\end{figure}
\newpage
\noindent
\underline{MODEL 3}\\\\
\begin{figure} 
\epsfxsize=8truecm
\centerline{\epsffile[50 263 506 693]{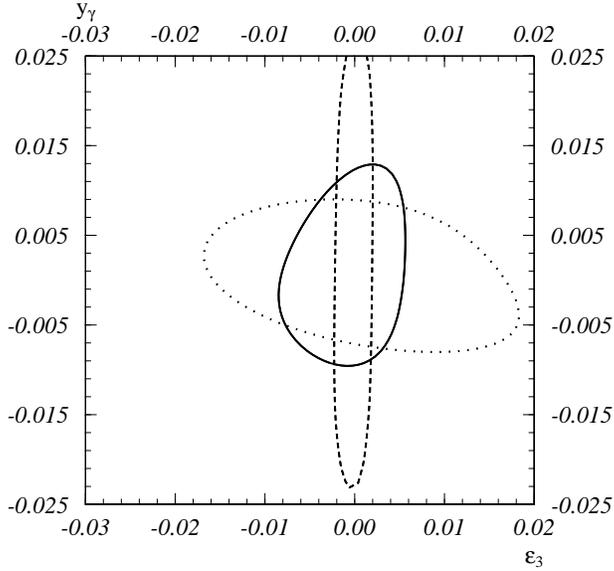}}
\caption{90\% C.L. bounds on the parameter space $(\epsilon_3,y_\gamma)$
in the case of model 1. The bounds are obtained from 
$\sigma_{\rm comb}$ at $E=500~GeV$ and $L=20~fb^{-1}$, and
 different degrees of polarization: the 
 dashed line corresponds to $P_e=0.9$,
 the continuous one to $P_e=0$,
 and the dotted one  to $P_e=-0.9$.
   }
\end{figure}
\begin{figure} 
\epsfxsize=8truecm
\centerline{\epsffile[50 263 506 693]{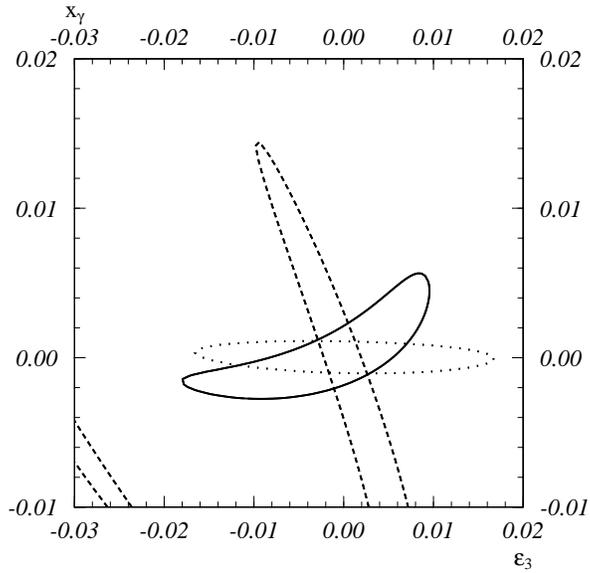}}
\caption{Same of Figure 8 for the parameter space $(\epsilon_3,x_\gamma)$
in the case of model 2.  }
\end{figure}
Let us now consider the so called BESS model (see ref. \cite{BESS}).
This is an effective model in which the existence of a new triplet of
vector fields ${\vec V}_\mu$ is assumed. These are the gauge fields
associated to a spontaneously broken local symmetry $SU(2)_V$. The new
vector particles mix with ${W}$ and $Z$. As a consequence at
energies well below their mass (assumed to be around the $TeV$ scale),
effective anomalous fermionic and trilinear couplings are generated
(see ref. \cite{lowen}). The parameters characterizing this model are the
gauge coupling of the ${\vec V}_\mu$, $g^{\prime\prime}$, and their
direct coupling to the fermions, $b$. Only $\epsilon_3$ and $\delta_z$
are different from zero and their are given by
\be
\epsilon_3=-\frac b 2+\left(\frac g {g^{\prime\prime}}\right)^2
\ee
\be
\delta_z=\frac{\kt}{2\s\kdt}\left(b-\frac 1{\kt^2}  
\left(\frac g {g^{\prime\prime}}\right)^2\right)
\ee

\begin{figure} 
\epsfxsize=8truecm
\centerline{\epsffile[50 263 506 693]{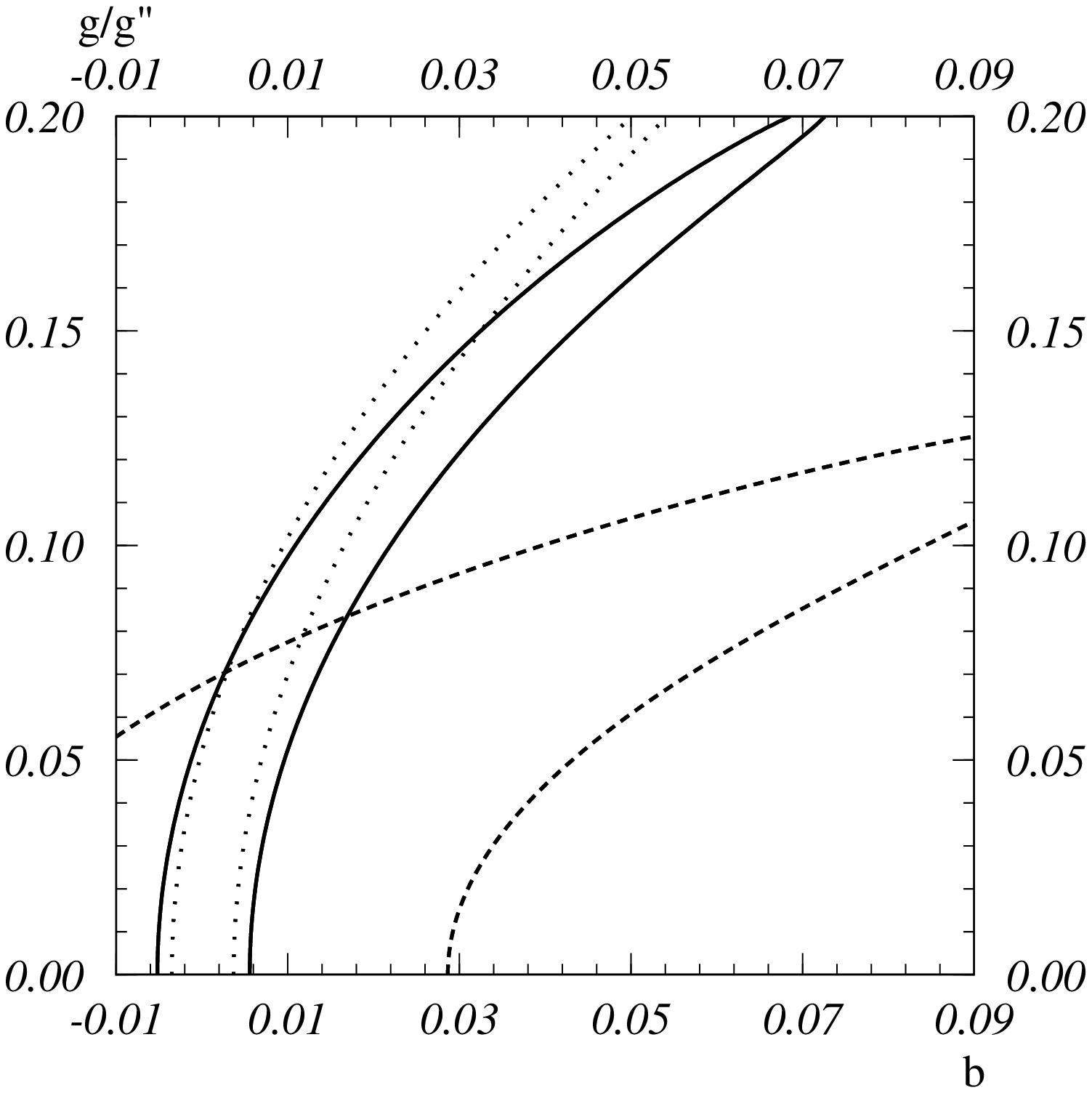}}
\caption{Same of Figure 8 for the parameter space $(b,g/g^{\prime\prime})$
in the case of  model 3 (BESS model).
  }\end{figure}

In Figs. 5 and 6 we give the 90\% C.L. bounds in the parameter space 
$(b, g/g^{\prime\prime})$ again at fixed energy and different observables
and at various energies respectively. Also in this case $\sigma_{LL}$
results to be the most restrictive observable. 
The BESS model is a generalization of a non-linear $\sigma$-model and 
it is not renormalizable. However at 1-loop level it is equivalent to a heavy
Higgs model with the Higgs mass playing the role of a cut-off.
Assuming 
the cut-off of the order of 1 $TeV$ we can evaluate 
$\epsilon_3^{\rm rad}$ with a Higgs mass equal to 1 $TeV$.
In Fig. 7 we give the
bounds obtained in this way 
on the parameter space arising from LEP1 experiments.
The
comparison with Fig. 6 shows the need of reaching an energy of at least
500 $GeV$ in order to get a real improvement.

Finally we have considered the possibility of  polarized electron beams.
In Figs. 8, 9 and 10 we show, for the previous models, the 
90\% C.L. bounds from $\sigma_{\rm comb}$ 
in the cases of $P_e=0,\pm 0.9$. We see that the general effect of
the polarization is to select different combinations of the anomalous parameters 
producing, as a consequence, a rotation of the allowed region. In this way, combining
polarized and non polarized differential cross-sections one is able to get significantly 
more restrictive bounds on the anomalous parameters. 

\resection{Conclusions}

The main point of this paper is the inclusion of  non-standard fermionic couplings
on the same footing of the anomalous gauge ones in the phenomenological
analysis of the $e^+e^-\to W^+W^-$ process for different options of the
next linear colliders. This
 is justified by the fact that new
physics will change generally both types of couplings, and also from the LEP1
results, that strongly encourage the hypothesis that, if they are there,
then they  should be of the same order of magnitude. The analysis shows
also that the these two different types of non-standard couplings are strongly
correlated. 

We have also studied the high-energy behaviour of the differential
cross-sections showing that there is a quadratic contribution
from the anomalous couplings which dominates the limit. Therefore
a linear approximation in the anomalous
terms is not justified, and, in fact,
 all our numerical analysis is done by using the exact
expressions.

In summary we have seen that the NLC (specially in the higher energy
option considered here) can give very stringent bounds on the anomalous
parameters also improving our knowledge of the fermionic couplings.
Furthermore, if  polarized electron beams will be
available, the bounds will increase dramatically.

\resection{Appendix A}

We give here the list of the coefficients $F^{(\alpha)}$ appearing in the 
high-energy expansion of the helicity amplitudes for $\alpha=1,1/2,0,-1/2,-1$.
We give separately the anomalous contributions and the SM ones. 

\begin{center}
{\bf{Longitudinal-Longitudinal Case}}
\end{center}
\bea
&& \delta F^{1/2,-1/2(1)}_{0,0}  = - \AAA\frac 1 {M_W^2}
   \Big ( \cot\theta  (\delta a_e^Z -  \delta b_e^Z) +\tan\theta
    (\delta_{z}  +
	x_{z} )-x_\gamma\Big )\\
&&F^{1/2,-1/2(0)}_{0,0}(SM)  = -\AAA \frac 1 {c_\theta^2}\\  
&&\delta F^{1/2,-1/2(0)}_{0,0}=-\AAA\Big (-\frac 2 {c_\theta^2}\frac{\delta M_W}
{M_W}+\frac 1 {\s\kt}(\delta a_e^Z-\delta 
b_e^Z)\nn\\
&&~~~~~~~~~~~~~~~~~~~~~ +\delta_Z\frac{\tan\theta}{\kt^2}-
x_z\tan\theta\frac{1-2\s^2}{\kt^2}+2 x_\gamma\Big  )\\
&&F^{1/2,-1/2(-1)}_{0,0}(SM)=-\AAA \frac {M_W^2}{\kt^4}\\
&&\delta 
F^{-1/2,1/2(1)}_{0,0} =\AAA\frac{1}{M_W^2}\Big (-\frac{2\sqrt{2}\delta 
a_W}{\s}-\cot\theta (\delta a_e^Z+\delta b_e^Z)\nn\\
&&~~~~~~~~~~~~~~~~~~~~~+\cot 2\theta 
(\delta_z+x_z)+x_\gamma\Big )\\
&&  F^{-1/2,1/2(0)}_{0,0}(SM)=-\frac {e^2} 2 \sin\phi\frac1{s^2_{2\theta}}\\
&&\delta F^{-1/2,1/2(0)}_{0,0}=\AAA\Big (-4\frac {\cos 2\theta}{\sin^2 2\theta}
\frac{\delta M_W}{M_W}-
\frac 1{\s\kt}(\delta a_e^Z+\delta b_e^Z)\nn\\
&&~~~~~~~~~~~~~~~~~~~~~-\frac{4\sqrt{2}}{\s}\delta 
a_W+\frac{\cot 2\theta}{\kt^2}(\delta_z-x_z c_{2\theta})-2 x_\gamma\Big )\\
&&F^{-1/2,1/2(-1)}_{0,0}(SM)=-\AAA\frac{M_W^2}{\s^2}\Big(-\frac{c_{2\theta} }{2
\kt^4}+2\frac{1-3\cos\phi}{1-\cos\phi}\Big)
\eea

\begin{center}
{\bf{Longitudinal-Transverse  Case}}
\end{center}
\bea
&&\delta F^{1/2,-1/2(1/2)}_{1,0}=\AAC \frac 1{M_W}\Big(2\cot\theta (\delta 
a_e^Z-\delta b_e^Z)\nn\\
&&~~~~~~~~~~~~~~~~~~~~~-(x_\gamma+y_\gamma)+\tan\theta(x_z+y_z+2\delta_z)\Big)\\
&&F^{1/2,-1/2(-1/2)}_{1,0}(SM)=\frac{e^2}{2\sqrt{2}}
(\cos\phi+1)\frac{M_W}{\kt^2}\\
&&\delta F^{-1/2,1/2(1/2)}_{1,0} =\AAB\frac 1{M_W}\Big(
-\frac{4\sqrt{2}}{\s}\delta a_W -2\cot\theta(\delta a_e^Z+\delta 
b_e^Z)\nn\\
&&~~~~~~~~~~~~~~~~~~~~~+(x_\gamma+y_\gamma)+\cot 2\theta(x_z+y_z+2\delta_z)\Big)\\
&& F^{-1/2,1/2(-1/2)}_{1,0}(SM)=-\frac{e^2}{\sqrt{2}}(\cos\phi-1)\Big(\cot 
2\theta+\frac 1{\s^2}\frac{\cos\phi}{1-\cos\phi}\Big) M_W\\
 &&\delta  F^{1/2,-1/2(1/2)}_{-1,0}=\AAB\frac 1{M_W}\Big(-2\cot\theta(\delta 
a_e^Z-\delta b_e^Z)\nn\\
&&~~~~~~~~~~~~~~~~~~~~~+ (x_\gamma+y_\gamma)-\tan\theta(x_z+y_z+2\delta_z)\Big)\\
&&  F^{1/2,-1/2(-1/2)}_{-1,0}(SM)=\frac{e^2}{2\sqrt{2}}(-\cos\phi+1)\frac 
{M_W}{\kt^2}\\
&&\delta  F^{-1/2,1/2(1/2)}_{-1,0}=\AAC\frac 
1{M_W}\Big(\frac{4\sqrt{2}}{\s}\delta a_W+2\cot\theta(\delta a_e^Z+\delta 
b_e^Z)\nn\\
&&~~~~~~~~~~~~~~~~~~~~~-(x_\gamma+y_\gamma)-\cot 2\theta(x_z+y_z+2 \delta_z)\Big)\\
&&F^{-1/2,1/2(-1/2)}_{-1,0}(SM)=-\AAC M_W \frac{1+2\kt^2}{\s^2\kt^2}
\eea
 
\begin{center}
{\bf{Transverse -Transverse  Case}}
\end{center}
\bea
&&\delta  F^{1/2,-1/2(1)}_{1,1}=\AAA\frac{1}{M_W^2}(y_\gamma-y_z\tan\theta)\\
&&F^{1/2,-1/2(0)}_{1,1}(SM)=0\\
&&\delta  F^{1/2,-1/2(0)}_{1,1}=-\AAA\Big(2\cot\theta(\delta a_e^Z-\delta 
b_e^Z)\nn\\
&&~~~~~~~~~~~~~~~~~~~~~+2 y_\gamma+2\delta_z\tan\theta- y_z\tan\theta\frac{ c_{2\theta}}
{c_\theta^2} \Big)\\
&&F^{1/2,-1/2(-1)}_{1,1}(SM)=-\frac{e^2} 2\sin\phi\frac{M_W^2}{\kt^2}\\
&&\delta  F^{-1/2,1/2(1)}_{1,1}=\AAA\frac{1}{M_W^2}(y_\gamma+y_z\cot 2 \theta)\\
&&F^{-1/2,1/2(0)}_{1,1}(SM)=0\\
&&\delta  
F^{-1/2,1/2(0)}_{1,1}=\AAA\Big(-\frac{4\sqrt{2}}{\s}\delta 
a_W-2\cot\theta(\delta a_e^Z+\delta b_e^Z)\nn\\
&&~~~~~~~~~~~~~~~~~~~~~+2\delta_z\cot 2\theta-2 
y_\gamma-y_z\cot 2\theta\frac{c_{2\theta}}{\kt^2}\Big)\\
&&F^{-1/2,1/2(-1)}_{1,1}(SM)=\AAA 
M_W^2\Big(-\frac1{\s^2\kt^2}+\frac{1}{\s^2}\frac{2\cos\phi}{1-\cos\phi}\Big)\\
 &&\delta  F^{-1/2,1/2(1)}_{1,-1}=0\\
&&F^{-1/2,1/2(0)}_{1,-1}(SM)=\AAA\frac 1{\s^2}\\
&&\delta  F^{-1/2,1/2(0)}_{1,-1}=e^2\sin\phi\frac{\sqrt{2}}{\s}\delta a_W\\
&&F^{-1/2,1/2(-1)}_{1,-1}(SM)=\frac{e^2} 2\sin\phi\frac{M_W^2}{\s^2}\\
&&\delta  F^{-1/2,1/2(1)}_{-1,1}=0\\
&&F^{-1/2,1/2(0)}_{-1,1}(SM)=-\AAA\frac 1{\s^2}\frac{1+\cos\phi}{1-\cos\phi}\\
&&\delta  
F^{-1/2,1/2(0)}_{-1,1}=-e^2\sin\phi\frac{\sqrt{2}}{\s}\frac{1+\cos\phi}
{1-\cos\phi}\delta a_W\\
 && F^{-1/2,1/2(-1)}_{-1,1}(SM)=-\frac{e^2} 2
\sin\phi \frac {M_W^2}{\s^2}\frac{1+\cos\phi}{1-\cos\phi}
\eea 

\begin{center}
\begin{bf} 
 ACKNOWLEDGMENTS
  \end{bf}
\end{center}

This work is part of the EEC Project "Tests of electroweak symmetry breaking 
and future European colliders", No. CHRXCT94/0579.

\end{document}